\documentclass[aps,prb,twocolumn,groupedaddress,showpacs]{revtex4}
\usepackage{graphicx}
\newcommand{\chem}[1]{\ensuremath{\mathrm{#1}}}
\def\func#1{\mathop{\rm #1}}

\begin{document}


\title{Berry's phase contribution to the anomalous Hall effect of gadolinium}


\author{S. A. Baily}
\email[]{sbaily@unm.edu}
\altaffiliation{Current address: Air Force Research Laboratory, Kirtland AFB, NM 87117}
\author{M. B. Salamon}
\affiliation{Department of Physics, University of Illinois Urbana-Champaign, Urbana, IL 61801}


\date{\today}

\begin{abstract}
When conduction electrons are forced to follow the local spin texture, the resulting Berry phase can induce an anomalous Hall effect (AHE). In gadolinium, as in double-exchange magnets, the exchange interaction is mediated by the conduction electrons and the AHE may therefore resemble that of \chem{CrO_2} and other metallic double-exchange ferromagnets. The Hall resistivity, magnetoresistance, and magnetization of single crystal gadolinium were measured in fields up to 30~T. Measurements between 2~K and 400~K are consistent with previously reported data. A scaling analysis for the Hall resistivity as a function of the magnetization suggests the presence of a Berry's-phase contribution to the anomalous Hall effect.
\end{abstract}


\maketitle

\section{introduction}
While many theories account for an anomalous Hall effect (AHE),
proportional to the magnetization of a material, these theories often
predict effects significantly smaller than those found in ferromagnetic
materials.\cite{PNDheer67,HRLeribaux66,RWKlaffky74,LBerger70,CLChien80,FEMaranzana67,JYe99}
An even more significant deficiency of the conventional theories
is that they predict an anomalous Hall resistivity that is proportional to a
power of the longitudinal resistivity, and in the absence of a metal insulator transition
cannot account for an AHE that peaks near the Curie temperature, \chem{T_C}. Recent
models based on a geometric, or Berry, phase have had
great success in describing the AHE in double-exchange
systems (e.g., manganites and chromium dioxide) and pyrochlores.\cite{JYe99,YLyandaGeller99,YLyandaGeller2001,HYanagihara2002,SHChun99,SHChun2000,SHChun2000b,YTaguchi2003,SOnoda2003}

The anomalous Hall effect in chromium dioxide, a metallic double-exchange ferromagnet,\cite{MAKorotin98} was
shown\cite{HYanagihara2002} to agree well with the description based on geometric phase first suggested by Ye, et al.\cite{JYe99}
In gadolinium, as in double-exchange magnets, the exchange interaction among localized (4\textit{f}) core spins is mediated by the
conduction electrons. The anomalous Hall effect may therefore resemble
that of \chem{CrO_2} and other metallic double-exchange ferromagnets. Monte-Carlo
simulations predict that the same spin-texture excitations that cause the anomalous
Hall effect in double-exchange systems are also intrinsic to Heisenberg ferromagnets.\cite{MJCalderon2001}
Thus it is reasonable to seek to explain the anomalous Hall effect in other systems
using the same theory.

Gadolinium has a unexpectedly large anomalous Hall effect.\cite{MChristen79} In particular,
when the applied magnetic field is parallel to the \textit{c}-axis the anomalous Hall resistivity peaks at $\rho_{xy}\approx -6~\mu\mathrm{\Omega~cm}$ just below \chem{T_C}.\cite{RSLee67} This
makes it a good candidate for showing a maximum near 2/3 of its saturation magnetization
as chromium dioxide does. Since gadolinium is metallic even above \chem{T_C}, conventional
theories cannot explain a maximum in the Hall effect near the transition temperature.
\section{sample preparation}
A \textit{c}-axis oriented gadolinium (99.99\% purity) single crystal was purchased from MaTecK GmbH.  Two cuts were made parallel to an in-plane axis direction, the sides were polished lightly to clean up rough edges from the saw cuts, and the \textit{c}-axis plane was thinned as much as possible. The resulting shape is a rectangular prism with an approximately square cross-section, and irregular ends. Gold contact pads were sputtered onto the sides of the sample.
\section{experimental results}
\begin{figure}
\includegraphics[width=8.5 cm,clip]{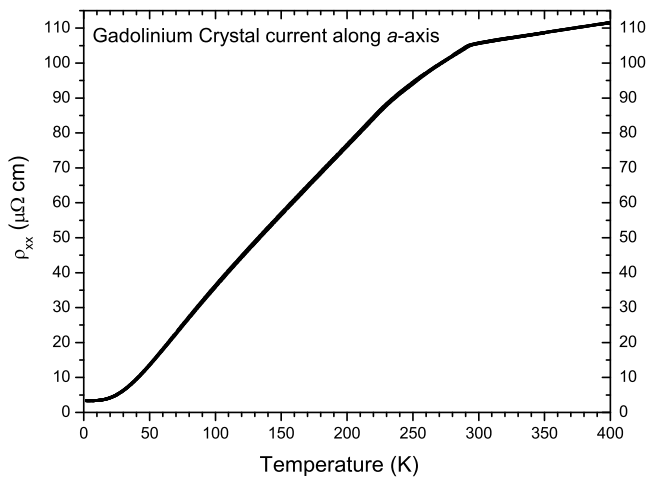}
\caption{Gadolinium resistivity vs.\ temperature.\label{gdRvsT}}
\end{figure}

\begin{figure}
\includegraphics[width=8.5 cm,clip]{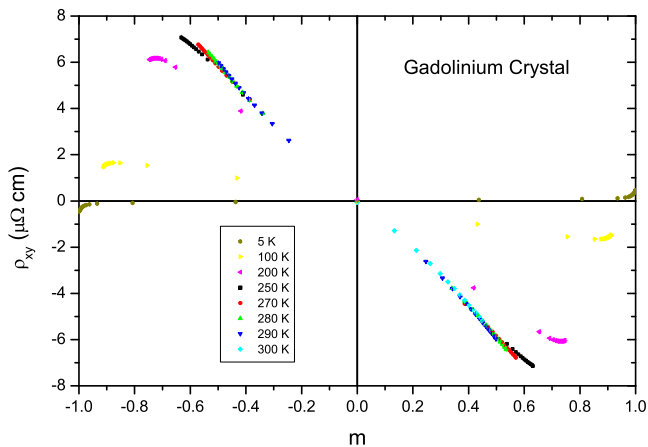}
\caption{(Color online) Gadolinium Hall resistivity vs.\ reduced magnetization.\label{gd7Trhoxyvtm}}
\end{figure}

Data were taken using a Quantum Design Physical Property Measurement System
(PPMS) in fields up to 7 T. The zero field resistivity for the gadolinium
crystal is shown in Fig.~\ref{gdRvsT}.
An alternating current (37~Hz) is applied along the \emph{a}-axis. An abrupt change in slope occurs at the ferromagnetic transition temperature. The residual resistivity ratio ($R_{300~\mathrm{K}}/R_{4.2~\mathrm{K}}$) is 31. For Hall effect and magnetization measurements, the field was applied along the \emph{c}-axis. The demagnetizing factor $N=0.5$, and the saturation magnetization is 7.7 $\mu_{B}/$Gd. The large values of the Hall conductivity and the magnetization allowed for very precise measurements. Fig.~\ref{gd7Trhoxyvtm} shows the Hall resistivity plotted vs.\ reduced magnetization ($m=M/M_{saturation}$); these data were collected in fields up to 7~T.
The Hall resistivity increases rapidly with magnetization below the Curie
temperature as domains are swept out. There is some indication that the data
maximize at $|\rho _{xy}|\approx 7~\mu\Omega$~cm when $|m|\approx
0.7$. It is conventional to separate the Hall resistivity into ordinary (OHE) and anomalous (AHE) contributions:
$\rho _{xy}=R_oB_{in}+R_s\mu _0M$,
where $B_{in}=\mu_0H_{applied}+\mu _0(1-N)M$. $R_o$ and $R_s$ are the ordinary and anomalous (or spontaneous) Hall coefficients respectively.
The upturns at large values of $m$ are from the OHE, which is small, but not completely negligible.

\begin{figure}
\includegraphics[width=8.5 cm,clip]{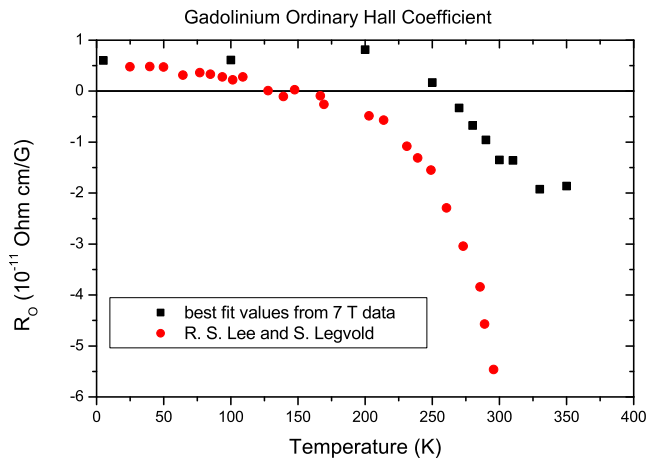}
\caption{(Color online) Gadolinium ordinary Hall coefficient vs.\ temperature. This represents the best fit to the data in fields below 7~T. R. S. Lee and S. Legvold's data are shown for comparison.\cite{RSLee67}\label{gdRovsT}}
\end{figure}

\begin{figure}
\includegraphics[width=8.5 cm,clip]{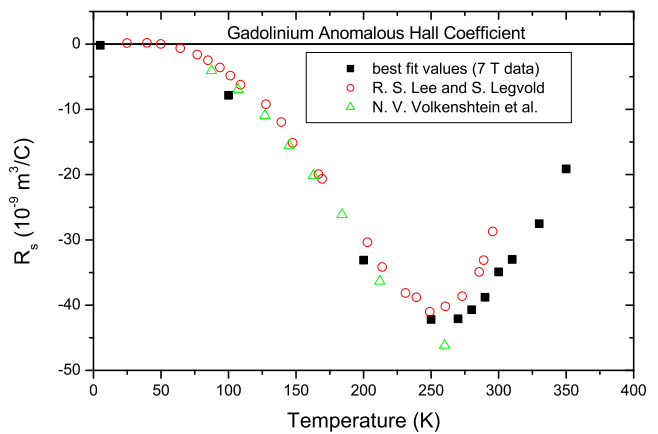}
\caption{(Color online) Gadolinium anomalous Hall coefficient vs.\ temperature. Previously reported data are shown for comparison.\cite{RSLee67,NVVolkenshtein66}\label{gdRsvsT}}
\end{figure}

It is difficult to make a reliable separation of the OHE and AHE contributions. To
obtain the values shown in Fig.~\ref{gdRovsT},
we first choose the anomalous Hall coefficient, $R_s$. Next, the corresponding term (linearly proportional to magnetization) is subtracted from the dataset. Then, a linear least squares fit of Hall resistivity versus internal field is made. The value chosen for the anomalous Hall coefficient is adjusted until the fitting error is minimized. The best-fit anomalous Hall coefficients are shown in Fig.~\ref{gdRsvsT}.
This method works even slightly above \chem{T_C}, because of the large demagnetizing correction; At temperatures significantly above \chem{T_C} the magnetization curves become linear in field, and this method fails. The other disadvantage to this method is that the Berry's phase theories predict that the anomalous Hall resistivity is linear in magnetization near \chem{T_C} only for low values of $m$.

The low temperature ordinary Hall coefficient agrees with previously reported values (see Fig.~\ref{gdRovsT}).\cite{RSLee67} The qualitative behavior is also similar. R. S. Lee and S. Legvold report that the ordinary Hall coefficient of gadolinium has temperature dependence which differs dramatically from those of lutetium and yttrium and cannot be explained by a two-band model.\cite{RSLee67}  They obtained a Hall coefficient which changes sign near 130~K (instead of 260~K, as seen in Fig.~\ref{gdRovsT}) and decreases even more rapidly as \chem{T_C} is approached. The most likely cause of these discrepancies is a problem with the separation of OHE and AHE. Lee and Legvold only applied 3~T, whereas the values reported here include data up to 7~T. Indeed, when a subtraction was attempted using the noisier 30~T data (see below), the ordinary Hall coefficient did not appear to change sign until \chem{T_C}. There are two possible explanations for this behavior. The simplest is that the AHE is underestimated, and the residual gives an apparent contribution to the OHE. The other possibility is that the sign change and the sharp increase in the magnitude of the Hall coefficient are real effects (possibly due to exchange splitting of the conduction band). In this case, the decreasing magnitude that we observe at higher fields and higher magnetization may be the result of an anomalous Hall effect that is not strictly linear in magnetization at high fields. This non-linearity, if real, would support the hypothesis that Berry's phase effects contribute to the anomalous Hall effect in gadolinium.  This contribution would decrease as the magnetization increases, thus giving rise to the apparent field dependence of the ordinary Hall coefficient. The ordinary Hall resistivity will not be subtracted from plots of the data because of this dilemma.

If the anomalous Hall effect results from the thermal excitation of
topological excitations, it is possible to use scaling relations for the
magnetization and expected Skyrmion density to obtain:\cite{HYanagihara2002}
\begin{equation}
\rho_{xy}^{A}T=\rho_{xy}^{0}T_{C} m[1-\func{D}(x)m^{(1-\alpha)/\beta}],
\label{skyrmion}
\end{equation}
where $\func{D}(x)$ is a scaling function of the scaling variable $x=t/h^{1/(\beta\delta)}$,
and $t$ and $h$ are the reduced temperature and magnetic field respectively. Along the critical
isotherm $t\equiv 1-T/T_{C}=0$, making $\rho _{xy}$ a function of $m$ only.

\begin{figure}
\includegraphics[width=8.5 cm,clip]{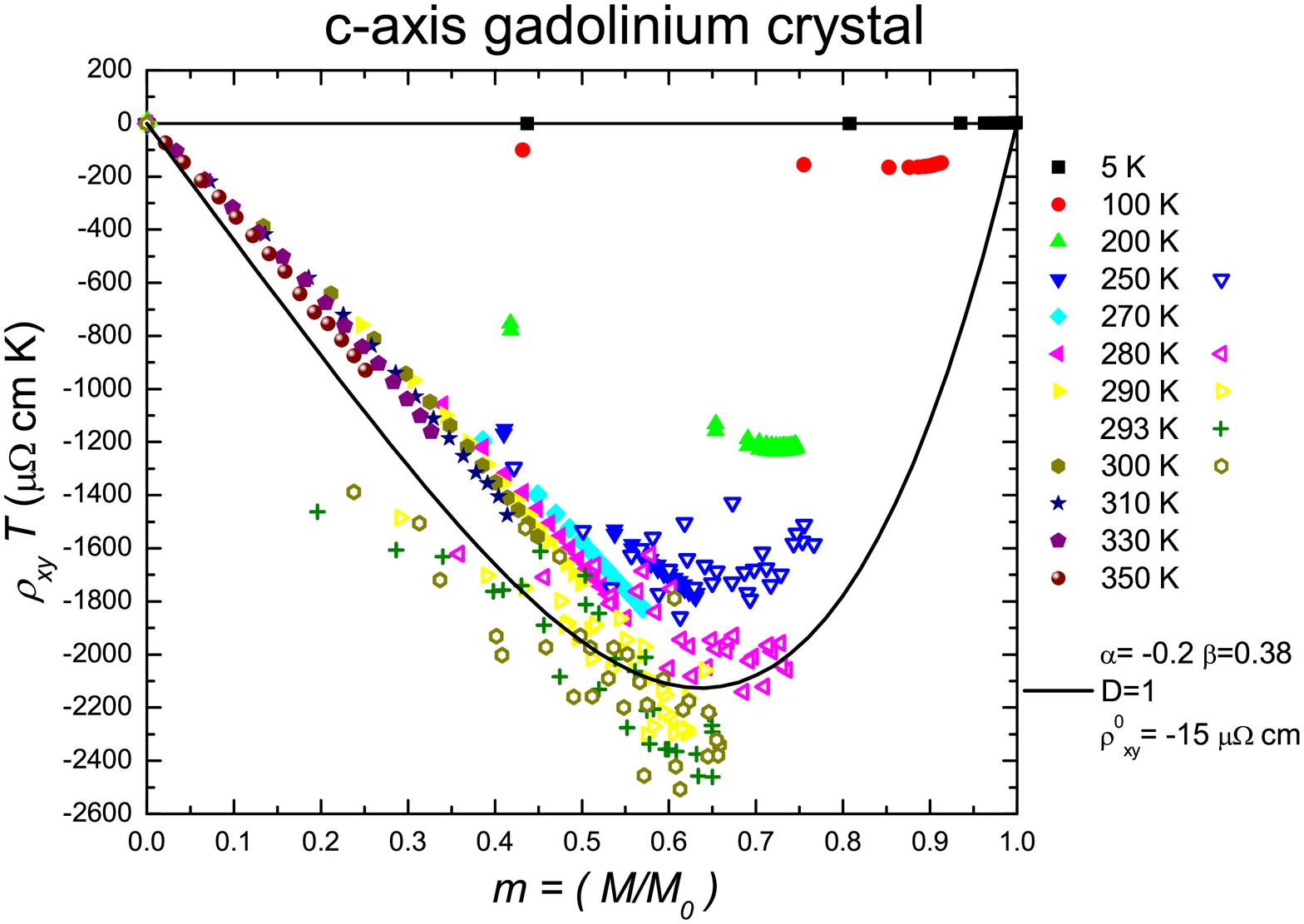}
\caption{(Color online) Temperature scaled Hall resistivity vs.\ reduced magnetization.\label{gdB2}}
\end{figure}

In an effort to extend the results in the vicinity of the Curie temperature to larger values
of $m$, we measured both the Hall resistivity and magnetization at the
National High Magnetic Field Laboratory in fields up to 30~T. The high
field data are consistent with those taken in the PPMS, but are noisier due to
problems both with the vibrating sample magnetometer and with pick-up
from ripple in the Bitter magnets. Nonetheless, there is a clear tendency
for the Hall resisitivity to reach an extremal value close to $m=2/3$. This is shown in
Fig.~\ref{gdB2}, where the closed symbols in the legend are from
the PPMS measurements and the remainder from the 30~T experiment. The solid line is the
Skyrmion expression (Eq.~\ref{skyrmion}) using the
critical exponents for gadolinium\cite{DSSimons74} and $\func{D}(0)=1$.

\section{Discussion}
Clearly the data in Fig.~\ref{gdB2} do not collapse well, yet suggest a
tendency to fit the Skyrmion picture. The initial slope at \chem{T_C},
$\rho_{xy}^{0}=-15~\mu\mathrm{\Omega~cm}$, depends on the Skyrmion density and
spin-orbit constant through:
\begin{equation}
\rho _{xy}^{0}=-\frac{1}{ne}\frac{\Phi _{0}}{\pi }\frac{\lambda _{so} n_{e} a S}{%
k_{B}T_{C}}\left\langle n\right\rangle .  \label{rhoxy0}
\end{equation}
Assuming $n_{e}=1$ carrier per Gd atom, $S=7/2$, and a Skyrmion density
$\left\langle n\right\rangle \approx 0.05$ near $T_{c}$, we estimate a
spin-orbit coupling constant of $\lambda _{so}\approx 12$~K. Although the fit is
consistent with the data, the data collapse is not so good. The spin-orbit coupling constant also seems rather large.
We can make a rough estimate of the spin-orbit coupling energy, as J. Ye et al.\ \cite{JYe99} have done for manganite, from the Hamiltonian:
\begin{equation}
H_{so}= -\frac{\vec{S}\cdot \vec{L}}{2m^2c^2}\frac{1}{r}\frac{\partial V}{\partial r}.
\end{equation}
Next we approximate the gradient of the potential using:
\begin{equation}
\frac{\partial}{\partial r}V=-\frac{\partial}{\partial r}\frac{Ze^2}{r} \approx \frac{Ze^2}{r_{d}a}
\end{equation}
where $r_{d}$ is the orbital radius, and $a$ is the lattice constant. Then an approximation of the spin-orbit coupling, $\lambda_{so}$, is given by:
\begin{equation}
\lambda_{so}=\frac{Ze^{2}\hbar^{2} k_{F_{z}}}{4m^{2}c^{2}r_{d}a}.
\end{equation}
In the free electron model:
\begin{equation}
k_{F_{z}}=\frac{\sqrt[3]{3\pi ^{2}}}{a\sqrt{3}},
\end{equation}
so
\begin{equation}
\lambda_{so} \approx 1.8 \left(\frac{Ze^{2}}{2mc^{2}r_{d}}\right)\left(\frac{\hbar^{2}}{2ma^{2}}\right).
\end{equation}
J. Ye et al.\ called the middle term the ``dimensionless coupling constant appropriate for \emph{d}-orbitals,'' and the final term the ``band kinetic energy.''\cite{JYe99} This rough estimate of the spin-orbit coupling constant works out to be about 9~K for gadolinium.

\begin{figure}
\includegraphics[width=8.5 cm,clip]{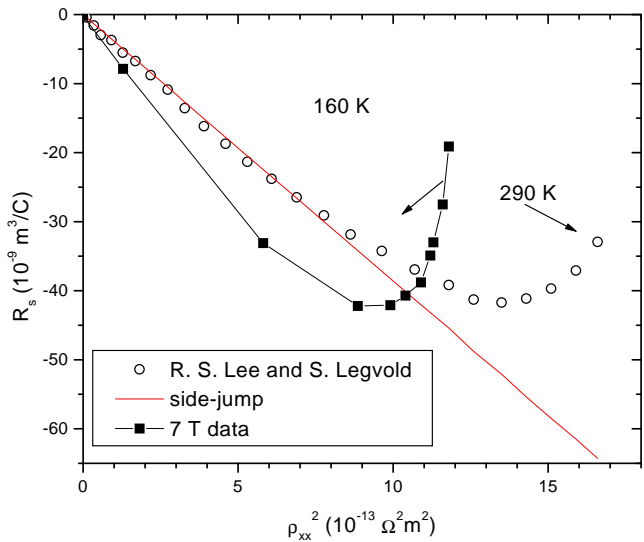}
\caption{(Color online) Anomalous Hall coefficient vs.\ resistivity squared. The residual resistivity has been subtracted. The ordinary Hall coefficient has been neglected when converting R. S. Lee and S. Legvold's data from $R_1$ to $R_s$.\protect\cite{RSLee67} The discrepancy in the plots is either due to an error in estimating the length between voltage contacts, or a systematic error in reading R. S. Lee and S. Legvold's data from their log-scale plot. The line is the side-jump prediction using the experimental coefficient for iron.\cite{RWKlaffky74,CLChien80,PNDheer67}\label{gdRsvsrho_xx2}}
\end{figure}

Unlike \chem{CrO_2}, where only those electrons participating in the
double-exchange contribute to the conductivity, Gd has both \emph{s} and
\emph{d}-electron contributions. It is not surprising, therefore, that the temperature
dependence (below 160~K) appears to be dominated by side-jump processes
($R_{s}\propto \rho _{xx}^{2}$),\cite{RSLee67} as seen in a plot of $R_{s}$
vs.\ $\rho _{xx}^{2}$ in Fig.~\ref{gdRsvsrho_xx2}. A side-jump
contribution, presumably from those portions of the Fermi surface that are
not strongly spin-polarized, should be distinguishable from the Skyrmion
contributions, for which $R_{s}\propto e^{-E_{c}/(k_{B}T)}/(k_{B}T)$.\cite{JYe99,HYanagihara2002}
As a further complication, R. S. Lee and S. Legvold's data show a low
temperature sign change of the anomalous Hall coefficient at a temperature different from the temperature at which the anomalous Hall coefficient changes sign; neither side-jump nor Skyrmion models can account for this. Extrapolation of the contribution
proportional to the square of the resistivity predicts a much larger Hall
effect above 200~K than is observed.
The ordinary Hall effect has been neglected when converting R. S. Lee and S. Legvold's data.\cite{RSLee67} Berger's prediction for the side-jump contribution is independent of the potential, so it should be essentially material independent, except for the enhancement due to band effects.\cite{LBerger70} Using the rough estimate calculated for iron (see Appendix~\ref{skewside}) gives a slope that is an order of magnitude too small for both iron and gadolinium.\cite{CLChien80} The straight line in Fig.~\ref{gdRsvsrho_xx2} has a coefficient that is one order of magnitude larger than this estimate. This coefficient is consistent in magnitude with experimental values for iron between 80~K and 267~K.\cite{CLChien80,RWKlaffky74,PNDheer67} While this term fits the lower temperature data, it is clearly too large near \chem{T_C}.

\bigskip We next explore whether the anomalous Hall effect exhibits better
scaling behavior if a side-jump contribution is removed. We assume,
arbitrarily, that a small side-jump process contributes to one sixth of the Hall effect at \chem{T_C}, i.e.,
\begin{equation}
R_s^{sj}=\frac{-254~\mathrm{\Omega^{-1}{cm}^{-1}}\times\rho_{xx}^{2}}{\mu_{0}M_{0}},
\end{equation}
and that the remainder is due to Berry's-phase processes.
The subtracted term is strictly linear in the magnetization, with a temperature dependence that depends on the square
of the zero field resistivity (excluding the residual resistivity). This side-jump contribution is of the same order of magnitude as expected theoretically, as shown in the appendix.  Figure~\ref{gdB2sideguess} is a plot of the difference versus reduced
magnetization, showing a better collapse of the data at both low and high fields, with an extremum in the vicinity of $m=0.6$.
\begin{figure}
\includegraphics[width=8.5 cm,clip]{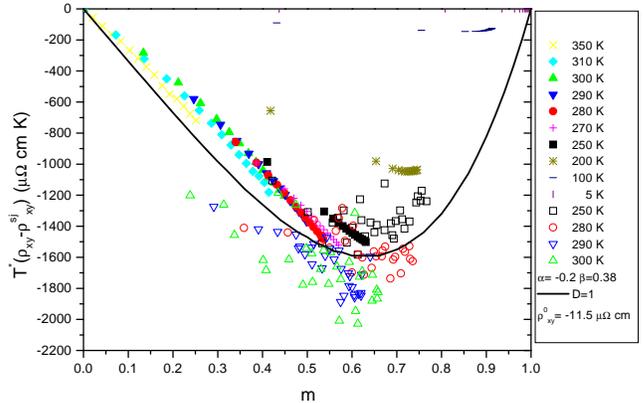}
\caption{(Color online) Possible Berry's phase contribution vs.\ reduced magnetization. The estimated side-jump contribution accounts for one third of the Hall effect at \chem{T_C}.\label{gdB2sideguess}}
\end{figure}
The evidence for a decrease in the anomalous Hall effect at high fields is even more convincing after subtracting the conventional term.  The line shown in Fig.~\ref{gdB2sideguess} is the same as in Fig.~\ref{gdB2}, except the initial slope is reduced, and $\func{D}(0)=1$ has been chosen. This same value for $\func{D}(x)$ also provided a good fit for \chem{CrO_2}.\cite{HYanagihara2002} In this case the spin-orbit coupling constant required would be 9~K, in close agreement with our rough estimate.
\appendix*
\section{Skew-scattering and Side-jump\label{skewside}}
More conventional explanations for the anomalous Hall effect include
side-jump and skew scattering.\cite{CMHurd72} Side-jump scattering is when
carriers scatter asymmetrically off impurities. Skew scattering
is a process caused by interference between spin-orbit coupling and second
order spin-flip scattering.\cite{JYe99} In conventional ferromagnets, this
theory yields values of $R_s$ two orders of magnitude smaller than
experimental data (according to some authors).\cite{JYe99,FEMaranzana67} Since the
carrier-electron spins must align with the localized core spins in double exchange
systems, spin-flip scattering cannot occur, and therefore skew scattering
cannot explain the Hall effect in manganites and other systems with strong
double exchange.

Karplus and Luttinger developed an early model for the anomalous Hall effect resulting from the spin-orbit interaction of spin-polarized conduction electrons.\cite{RKarplus54} Their model gave $R_s\propto\rho_{xx}^2$, but Smit criticized their model arguing that a periodic potential could not cause scattering and produce the anomalous Hall effect.\cite{JSmit55} Smit's theory, known as skew scattering, is based on anisotropic scattering caused by the spin-orbit interaction .\cite{JSmit58} After scattering off of an impurity, the momentum of the charge carriers is changed.  Spin-orbit coupling makes scattering to one side more likely; this gives rise to the Hall effect. Skew scattering is generally distinguished by $R_s\propto\rho_{xx}$,\cite{CLChien80} but can also give terms proportional to the square of the resistivity. The quadratic term occurs at high impurity concentrations (simultaneous scattering from multiple impurities) and from phonon scattering (at least above the Debye temperature).\cite{JSmit58} Leribaux \cite{HRLeribaux66,CLChien80} estimates the phonon scattering contribution in iron as:
\begin{equation}\label{skewscattering}
R_s=\frac{20.9~\mathrm{\Omega^{-1}{cm}^{-1}}}{\mu_0 M_s(T)}\rho_{xx}^2[1+T^2\times 1.12\times 10^{-8}~\mathrm{K^{-2}}].
\end{equation}

Somewhat later, Berger proposed the side-jump mechanism that yields $R_s\propto\rho_{xx}^2$.\cite{LBerger70} The side-jump mechanism occurs when the center of mass of a carrier's wave packet is translated to the side while inside the scattering potential. The effect can be envisioned by picturing light striking a window at an angle. The refractive index of the window results in a displacement of the light's path, but no change in direction because both glass/air interfaces are parallel. In general, this translation can be in any direction, but only asymmetric (due to the spin-orbit interaction) sideways jumps will directly contribute to the Hall effect. Klaffky and Coleman \cite{RWKlaffky74,CLChien80} estimate the side-jump scattering contribution in iron to be 5 times larger than the skew scattering contribution (Eq.~(\ref{skewscattering})), and given by:
\begin{equation}
R_{s}^{sj}=\frac{100~\mathrm{\Omega^{-1}{cm}^{-1}}}{\mu_0 M_s}\rho_{xx}^2.
\end{equation}

Recently, the skew and side-jump mechanisms have been treated simultaneously using a model based on the Kubo formalism and the Dirac equation.\cite{ACrepieux2001} Experimental results for single crystal iron show that the anomalous Hall coefficient is proportional to the square of the resistivity between 75~K and room temperature.\cite{PNDheer67,RWKlaffky74,CLChien80} The experimental coefficients, which range from $9.3\times 10^2~\mathrm{\Omega^{-1}{cm}^{-1}}$ to $1.44\times 10^3~\mathrm{\Omega^{-1}{cm}^{-1}}$, are much larger than either estimate.\cite{PNDheer67,RWKlaffky74,CLChien80} These results do not conclusively eliminate these mechanisms as the major source of the anomalous Hall effect, because the estimates are only valid to about one order of magnitude.

\begin{acknowledgments}
This material is based upon work supported by the U.S. Department of Energy, Division of Materials Sciences under Award No.\ DEFG02-91ER45439, through the Frederick Seitz Materials Research Laboratory at the University of Illinois at Urbana-Champaign. The authors would like to thank the National High Magnetic Field Laboratory in Tallahassee for the use of their facilities. This paper was completed while an author, Scott Baily, held a National Research Council Research Associateship Award at Air Force Research Laboratory.
\end{acknowledgments}

\bibliography{thesisplus}

\end{document}